\documentclass[prl,aps,twocolumn,preprintnumbers, showpacs, nofootinbib,superscriptaddress,notitlepage]{revtex4-1}
\usepackage{amsmath,graphicx,epsfig,amssymb}
\usepackage{inputenc}
\usepackage[usenames]{color}
\usepackage{ulem} 
\usepackage{bigstrut}
\usepackage{slashed}
\usepackage{multirow}
\usepackage{dsfont}
\usepackage{xcolor}
\usepackage{soul}

\allowdisplaybreaks

\newcommand{\nn}{\nonumber}
\def\slash#1{#1 \hskip-0.45em /}

\begin{document}

\title{A realistic method to access  heavy meson light-cone distribution amplitudes from first-principle}

\author{Xue-Ying Han}
\affiliation{Institute of High Energy Physics, CAS, Beijing 100049, China}
\affiliation{School of Physics, University of Chinese Academy of Sciences, Beijing 100049, China}

\author{Jun Hua} 
\affiliation{Key Laboratory of Atomic and Subatomic Structure and Quantum Control (MOE), Guangdong Basic Research Center of Excellence for Structure and Fundamental Interactions of Matter, Institute of Quantum Matter, South China Normal University, Guangzhou 510006, China}
\affiliation{Guangdong-Hong Kong Joint Laboratory of Quantum Matter, Guangdong Provincial Key Laboratory of Nuclear Science, Southern Nuclear Science Computing Center, South China Normal University, Guangzhou 510006, China}

\author{Xiangdong Ji}
\affiliation{Department of Physics, University of Maryland, College Park, MD 20742, USA}

\author{Cai-Dian L\"u}
\affiliation{Institute of High Energy Physics, CAS, Beijing 100049, China}
\affiliation{School of Physics, University of Chinese Academy of Sciences, Beijing 100049, China}

\author{Wei Wang}
\email{Corresponding author: wei.wang@sjtu.edu.cn}
\affiliation{Shanghai Key Laboratory for Particle Physics and Cosmology, Key Laboratory for Particle Astrophysics and Cosmology (MOE), School of Physics and Astronomy, Shanghai Jiao Tong University, Shanghai 200240, China}

\affiliation{Southern Center for Nuclear-Science Theory (SCNT), Institute of Modern Physics, Chinese Academy of Sciences, Huizhou 516000, Guangdong Province, China}

\author{Ji Xu }
\email{Corresponding author: xuji\_phy@zzu.edu.cn}
\affiliation{School of Nuclear Science and Technology, Lanzhou University, Lanzhou 730000, China}

\author{Qi-An Zhang }
\email{Corresponding author: zhangqa@buaa.edu.cn}
\affiliation{School of Physics, Beihang University, Beijing 102206, China}

\author{Shuai Zhao}
\affiliation{Department of Physics, Tianjin University, Tianjin 300350, China}

\begin{abstract} 
Lightcone distribution amplitudes (LCDAs) of heavy meson within heavy quark effective theory (HQET) are  crucial for predicting  physical observables  in $B$ decays, but  unfortunately there is  no first-principle result  due to severe challenges. After analyzing these  challenges,  we propose a realistic  method to determine heavy meson LCDA. We  utilize  equal-time correlations and  incorporate a dynamic quark field for a fast moving heavy quark. To verify this method, we make use of lattice QCD simulation on a lattice ensemble with spacing $a = 0.05187$\,fm. The preliminary findings for HQET LCDAs qualitatively align with phenomenological models, and  the fitted result for the first inverse moment $\lambda_B^{-1}$ is consistent with the experimentally constrain from $B \to \gamma \ell\nu_\ell$. We explore how our findings can reduce model uncertainties in  predictions of heavy-to-light form factors at large recoil. These results demonstrate the promise of our method in providing first-principle predictions for heavy meson LCDAs.
\end{abstract}
\maketitle

{\it Introduction:} The investigation of heavy meson decays, exemplified by the $B$-meson, offers profound insights into the non-perturbative realm of QCD and provides a stringent test for the boundaries of the standard model. At the heart of this analysis lies the concept of the light-cone distribution amplitude (LCDA) for heavy mesons~\cite{Beneke:1999br,Beneke:2000ry,Keum:2000wi,Lu:2000em}. 
These LCDAs  are defined in heavy quark effective theory (HQET) and encode the information about the probabilities of finding the quark and antiquark carrying certain momentum inside heavy meson~\cite{Grozin:1996pq}.  In addition they are essential for understanding the dynamics of   strong force at the interface between the long-range hadronic properties and the short-range quark-gluon degrees of freedom.

Though the ultraviolet behavior is understandable   from  QCD perturbation theory~\cite{Lange:2003ff,Lee:2005gza,Kawamura:2008vq,Bell:2013tfa,Feldmann:2014ika,Braun:2019wyx}, establishing a reliable result on the full distribution of heavy meson LCDAs is not available yet.    Constructing models or parameterizations~\cite{Grozin:1996pq,Braun:2003wx,Beneke:2018wjp} involves assumptions about the internal structure of heavy mesons, and the choice of model will unavoidably  introduce   uncertainties and biases. Recent studies  have utilized these models in the framework of $B$ meson light-cone sum rules (LCSRs) to calculate the form factors for  $B\to K^*$  and $B\to \pi$~\cite{Gao:2019lta,Cui:2022zwm}.  The  obtained results at $q^2=0$  are as follows:
\begin{align}
{\cal V}_{B\to K^*}(0) =& 0.359^{+0.141}_{-0.085}\Big|_{\lambda_B}{}^{+0.019}_{-0.019}\Big|_{\sigma_1}{}^{+0.001}_{-0.062}\Big|_{\mu}\nonumber\\
& {}^{+0.010}_{-0.004}\Big|_{M^2} {}^{+0.016}_{-0.017}\Big|_{s_0}{}^{+0.153}_{-0.079}\Big|_{\varphi_{\pm}(\omega)},  \label{eq:ErrorsOfBtoKFF} \nonumber\\
 f_{B\to \pi}^{0}(0)=&0.122\times\bigg[ 
1\pm0.07\Big|_{S_0^\pi} {} \pm0.11\Big|_{\Lambda_q} \nonumber\\
& \quad \pm0.02\Big|_{\lambda_E^2/\lambda_H^2}
 {}^{+0.05}_{-0.06}\Big|_{{M^2}}  \pm0.05\Big|_{2\lambda_E^2+\lambda_H^2}\nonumber\\
&\quad ^{+0.06}_{-0.10}\Big|_{\mu_h } \pm0.04\Big|_{\mu} {}
^{+1.36}_{-0.56}\Big|_{\lambda_{B}}{}^{+0.25}_{-0.43}\Big|_{ \sigma_1, \sigma_2}\bigg],
\end{align}
which uncertainties with the subscript $\lambda_B$  and   $\sigma_{1,2}$ in these form factors  are from parametric errors in heavy meson LCDAs,  while the one with the ${\varphi_{\pm}(\omega)}$ is induced by varying models  for the LCDAs. It is evident that these uncertainties  are dominant.   Therefore, the  lack of precise knowledge on LCDAs of heavy mesons from first-principles has been  the long-standing primary obstacle in making reliable predictions for heavy meson decays.

In this Letter, we propose a novel and realistic method to determine heavy meson LCDAs from first-principles of QCD which can cure theoretical difficulties as we will explain in the following. Explicitly,  we propose a sequential effective field theory and make use of equal time correlations, also named as quasi distribution amplitudes (quasi-DAs), of heavy meson with a large momentum component $P^z$. By designing  a hierarchical ordering for the scales $P^z \gg m_H \gg \Lambda_{\rm{QCD}}$,  we point out that dynamics in these scales can be separated by integrating out the $P^z$ and $m_H$ in two steps, and obtain the required  LCDAs in HQET at the final step. To verify this proposal, we perform a lattice QCD simulation of quasi-DAs on a lattice ensemble with $a=0.05187$\,fm. Our preliminary  findings for HQET LCDAs qualitatively align with the existing phenomenological models.  Using a model-independent parametrization,  we also determine the first inverse moment and the obtained result $\lambda_B= 0.389(35)$\,GeV lies in the experimentally  constrained region $\lambda_B>0.24$\,GeV derived from $B \to \gamma \ell\nu_\ell$ measurement~\cite{Belle:2018jqd}. The overall agreement demonstrates the promise of our method in providing first-principle predictions for heavy meson LCDAs.

{\it Challenges in determination of heavy meson LCDA and the solution:}
The leading twist  LCDA for a heavy meson is defined as~\cite{Grozin:1996pq}:
\begin{align}
  &\varphi^+(\omega,\mu) =  \frac{1}{i \tilde{f}_{H}(\mu) m_H} \int_{-\infty}^{+\infty} \frac{dt}{2\pi} \, e^{i\omega n_+ \cdot v t} \nn\\
   &\quad \times \left\langle 0\left|\bar{q}(t n_+) \slash{n}_+\gamma_5 W_c(t n_+, 0) h_v(0)\right|  H(v)\right\rangle \,,  \label{eq:def_of_HQET_LCDA}
\end{align}
where $h_v$ denotes heavy quark field in HQET, with the velocity $v$ satisfying $v^2=1$. ${q}$ denotes the light quark field with momentum $\omega$. $t$ denotes the spatial separation between the heavy and light quark fields.  $W_c$ is a Wilson line connecting the heavy and light quark fields to ensure the gauge invariance.

The first-principle calculation of heavy meson LCDA from lattice QCD would encounter significant difficulties. 
First of all,  simulating a light-cone distribution on the lattice is a challenging endeavor, made even more complex when incorporating an effective heavy quark field like $h_v$.  More critically, the coexistence of a light-like Wilson line and the HQET field $h_v$ leads to cusp divergences that will forbid all traditional lattice calculations. This can be elucidated by presenting a perturbative one-loop result for the involved operators~\cite{Braun:2003wx}: 
\begin{align}
\label{eq:cusp}
& O_v^{P, \mathrm{ren}}(t,\mu) = O_v^{P, {\rm bare}}(t) 
 \nonumber\\
&\;\;\;\; + \frac{\alpha_sC_F}{4\pi} \bigg\{\left(\frac{4}{\hat \epsilon^2} + \frac{4}{\hat \epsilon}\ln (it\mu)\right) O_v^{P, {\rm bare}}(t)  \nonumber\\
&\;\;\;\; - \frac{4}{\hat \epsilon}\int_0^1 du \frac{u}{1-u}[O_v^{P, {\rm bare}}(ut)-O_v^{P, {\rm bare}}(t)  ] \bigg\},
\end{align}
with  $d=4-\epsilon$  and the standard notation $2/\hat \epsilon=2/\epsilon -\gamma_E +\ln (4\pi)$.   
In the local limit $t \to 0$, the $\ln(it\mu)$ term will diverge.
Thus, the operator product expansion (OPE) for heavy meson LCDAs is ineffective, leading to ambiguities in defining the non-negative moments of heavy meson LCDAs.   For these reasons, while there have been numerous lattice QCD calculations for light meson LCDAs—both moments~\cite{Braun:2006dg,Boyle:2006pw,Arthur:2010xf,Braun:2015axa,Braun:2016wnx,Bali:2017ude,RQCD:2019osh,RQCD:2019osh,Detmold:2021qln} and full distributions~\cite{Bali:2017gfr,Zhang:2017bzy,Hua:2020gnw,LatticeParton:2022zqc,Zhang:2017zfe,Zhang:2020gaj},  no result for heavy meson LCDAs is available since the proposal in Ref.~\cite{Grozin:1996pq}.


To circumvent the above challenges, two directions are in general possible. Using an off-lightcone Wilson line approach is one option~\cite{Kawamura:2018gqz, Wang:2019msf,Zhao:2020bsx,Xu:2022krn,Xu:2022guw,Hu:2023bba,Hu:2024ebp}, yet it encounters a significant signal-to-noise problem when dealing with the $h_v$ field on the lattice~\cite{Mandula:1990fit,Mandula:1993sj,Meinel:2010uji}. The other option is  to construct the LCDA using QCD heavy quark fields.  The utilization of heavy quark with finite masses serves as an analogy to critical phenomena. The effective heavy quark field is obtained with the $m_Q\to \infty$ limit, which  represents a fixed point, usually rendering direct studies of correlations impractical. Accessing properties on fixed point can involve examining correlation functions in its vicinity and charting a trajectory for exploration. Shifting from this fixed point corresponds to the finite heavy quark mass and allows a direct simulation.   It is important to emphasize that successfully implementing this approach is not trivial. Instead, it is crucial to ensure that a quantity is constructed with the same infrared behavior as the HQET LCDA we ultimately desire. 
This requires boosting the heavy meson and  defining an LCDA that includes only QCD fields: 
\begin{align}
  &\phi(y, \mu) =  \frac{1}{i f_{H}} \int_{-\infty}^{+\infty} \frac{d\tau}{2\pi} \, e^{i y P_H \tau n_+} \nn\\
  &\quad \times \left\langle 0\left|\bar{q}(\tau n_+) \slash{n}_+\gamma_5 W_c(\tau n_+, 0) Q(0)\right| H(P_H)\right\rangle \,. \label{eq:def_of_QCD_LCDA}
\end{align}

The same infrared structure ensures that $\phi$ and $\varphi^+$ can be connected through a factorization
 \begin{align}
	\varphi^+ (\omega, \mu)=\frac{1}{m_H} \frac{f_H}{\widetilde{f}_H} \frac{1}{\mathcal{J}_{\mathrm{peak}}}
	\phi(y, \mu) \,,\label{eq:peak_varphi_+}
\end{align}
with $\omega=ym_H$.
The heavy meson mass $m_H$ is treated as a high scale in this factorization, which satisfying $m_H\gg\Lambda_{\mathrm{QCD}}$. It will contribute to the hard-collinear modes of the QCD LCDA~\cite{Ishaq:2019dst,Zhao:2019elu,Beneke:2023nmj}, and be collected into the perturbative jet function~\cite{Beneke:2023nmj}
\begin{align}
	\mathcal{J}_{\text {peak }} =1&+\frac{\alpha_s C_F}{4\pi}\bigg( \frac{1}{2}\ln^2\frac{\mu^2}{m_H^2}+\frac{1}{2}\ln\frac{\mu^2}{m_H^2}+\frac{\pi^2}{12}+2\bigg) \nn\\
	&+\mathcal{O}\left(\alpha_s^2\right)\,. 
\end{align}

After surmounting the challenges posed by the HQET field $h_v$, the subsequent phase involves implementing the QCD LCDA on lattice QCD. Drawing inspiration from large momentum effective theory (LaMET)~\cite{Ji:2013dva,Ji:2014gla} (see Ref.~\cite{Ji:2020ect,Cichy:2018mum} for recent reviews), the QCD LCDA is linked to an equal-time correlator with a substantial momentum $P^z$, referred to as the quasi DA:
\begin{align}
\tilde{\phi}(x, P^z)=\int \frac{dz}{2\pi} e^{-ix P^z z}\left\langle 0\left|\bar{q}(z)\Gamma W_c(z,0) Q(0) \right|H(P^z)\right\rangle_R.
\label{eq:def_of_quasi_LCDA}
\end{align}
where the lower script ``R'' denotes the renormalized matrix element.
The matrix element contains equal-time correlated nonlocal operator can be directly simulated on lattice QCD \cite{LatticeParton:2024zko}.
As  the $P^z$ tends towards infinity, the quasi DA will progressively converge towards the QCD LCDA. They also exhibit the same infrared structure, allowing them to be connected through a factorization:
 \begin{align}
\tilde{\phi}(x,P^z) =& \int_0^1 C \left(x, y, \frac{\mu}{P^z}\right) \phi(y, \mu) \nonumber\\
 & +\mathcal{O} \left(\frac{m_H^2}{(P^{z})^2},\frac{\Lambda_{\mathrm{QCD}}^2}{(xP^z, \bar xP^z)^2}\right)\,.
\label{eq:LaMETmatching}
\end{align}
Therefore $P^z$ serves as the ultra-violet scale, satisfying $P^z\gg m_H, \Lambda_{\mathrm{QCD}}$.

\begin{figure*}[!th]
\centering
\includegraphics[width=0.8\textwidth]{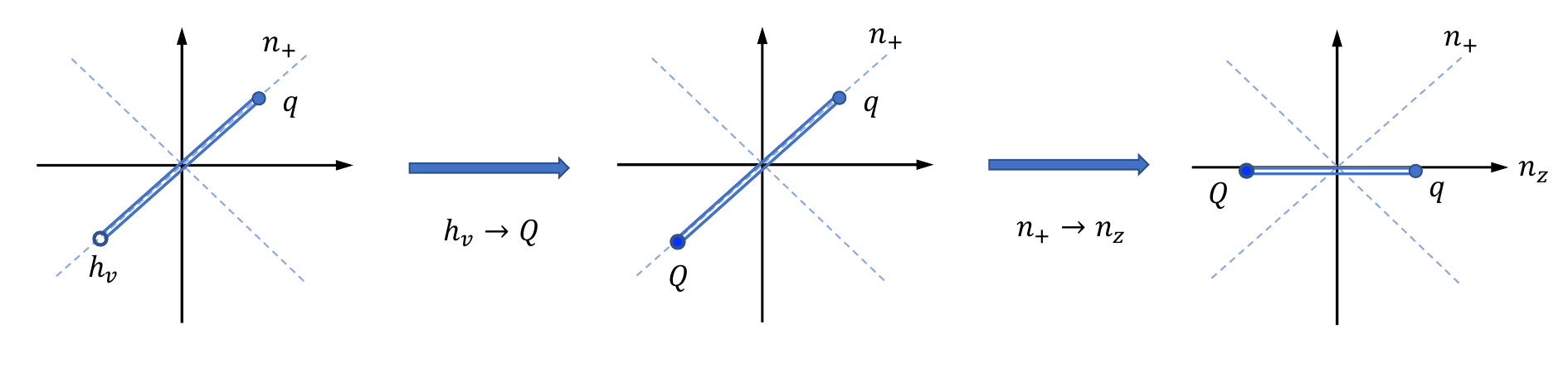}
\caption{Illustration of the two-step matching process to address issues in the first-principle calculation of heavy meson LCDA: emplying the dynamic heavy quark field can eliminate the cusp divergence, and establishing the equal-time correlation allows to circumvent the necessity of lightcone quantities.   }
\label{fig:2stepmatching}
\end{figure*}

Therefore,  the above discussions can be summarized as a two-step factorization scheme, as illustrated by Fig.~\ref{fig:2stepmatching}. 
To circumvent the challenges due to lattice simulations of HQET fields and the cusp divergence stemming from the nonlocal HQET operator matrix element, we first convert the HQET field into a QCD field, denoted by the first arrow in Fig.~\ref{fig:2stepmatching}. This process is implemented within boosted HQET factorization and establishes the hierarchy $m_H \gg \Lambda_{\mathrm{QCD}}$.  The QCD LCDA involves two distinct scales,$m_H$ and $\Lambda_{\mathrm{QCD}}$. However, as depicted by the second arrow in the figure, within the LaMET framework, one can derive the QCD LCDA via a lattice-computable equal-time correlated matrix element involving a highly boosted external state. To achieve this, $P^z$ must satisfy $P^z \gg m_H, \Lambda_{\mathrm{QCD}}$. Thus, in line with the essence of the two-step factorization, the hierarchical ordering of the three scales is $P^z \gg m_H \gg \Lambda_{\textrm{QCD}}$, imposing restrictions for the subsequent lattice validation.

{\it Lattice QCD verification:} 
It is crucial to emphasize that HQET LCDAs are not dependent on the heavy quark mass, which are due to heavy quark flavor symmetry. Hence,   one has the flexibility to select $m_Q$  as the charm quark mass $m_c$, the bottom quark mass $m_b$, or any other appropriate value, with the HQET LCDAs produced remaining unaffected. On the other hand, the heavy quark mass $m_Q$  needs to be carefully selected to alleviate the impact of power corrections originating from $m_H^2/(P^z)^2$  and $\Lambda_{\mathrm{QCD}}/m_H$  in the factorization processes. Our optimal choice is to concentrate on the charmed meson in this study.   

{To validate our method, we conduct a numerical simulation using the  gauge configurations produced by the CLQCD collaboration \cite{Hu:2023jet}, with $N_f=2+1$ flavor clover fermions and Symanzik gauge action. To guarantee the hierarchy among the three characteristic scales of quasi DA and minimizes the influence of power corrections in the two-step factorization process, we employ the finest ensemble at lattice spacing $a=0.05187$\,fm with sea quark masses corresponding to $m_{\pi}=317.2$\,MeV and $m_K=536.1$\,MeV. More details can be found in the supplemental paper \cite{LatticeParton:2024zko}.}

We perform numerical simulations of the quasi matrix elements of the $D$ meson, with a mass of $m_D \simeq 1.92$\,GeV, on 549 gauge configurations using boosted momenta $P^z$ up to $P^z \simeq 3.98$\,GeV. By utilizing the state-of-the-art self renormalization scheme \cite{LatticePartonLPC:2021gpi} and extrapolating the long-range correlations  through a physics-inspired extrapolation form \cite{Ji:2020brr}, we obtain the distributions of renormalized quasi matrix elements of heavy $D$ meson in coordinate space. 
 
\begin{figure}[!th]
\centering
    \includegraphics[width=0.45\textwidth]{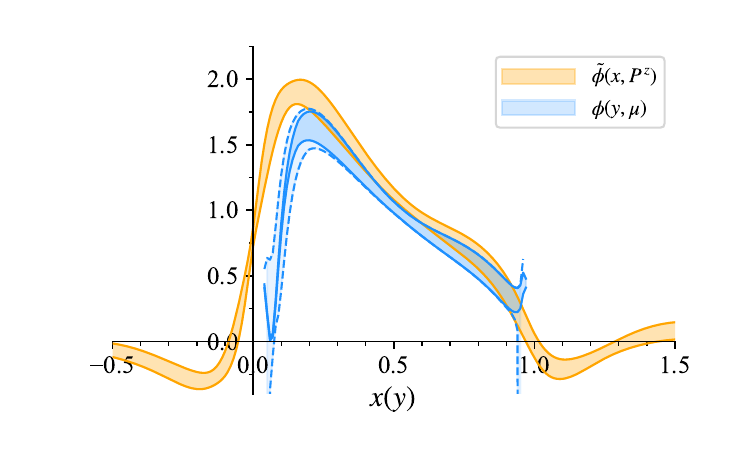}
\caption{Quasi DA $\tilde{\phi}(x,P^z)$ with $P^z=3.98$GeV, and matched QCD LCDA $\phi(y,\mu)$ from NLO kernel. The scale $\mu$ is chosen as $m_D$. The result of $\tilde{\phi}(x,P^z)$ only contain statistical error, while $\phi(y,\mu)$ contains both statistical error (solid lines in the blue band) and systematic one arising from the scale variation (dashed lines). }
\label{fig:Compare_mom_LCDA}
\end{figure}

As mentioned in Eq.(\ref{eq:def_of_quasi_LCDA}), the quasi DA $\tilde{\phi}(x,P^z)$ of heavy $D$ meson can be obtained from Fourier transforming its renormalized matrix elements into momentum space. The numerical results of $\tilde{\phi}(x,P^z)$ at $P^z=3.98$\,GeV is illustrated as the orange band in Fig.~\ref{fig:Compare_mom_LCDA}.

Applying the matching relation given in Eq.~(\ref{eq:LaMETmatching}), one can establish the connection between quasi DA $\tilde{\phi}(x, P^z)$ and QCD LCDA $\phi(y,\mu)$ with the  matching kernel  at next-to-leading order in $\alpha_s$ given in Ref.~\cite{Liu:2018tox,LatticeParton:2024zko}. 
In practice, we employ the renormalization group (RG) resummation starting from the lattice scale $\mu_0$ to $\mu=m_D$, which satisfies both the factorization scale of $\tilde{\phi}(x,P^z)$, and the renormalization scale of $\phi(y,\mu)$. The numerical result of $\phi(y, \mu)$ is shown as the blue band in Fig.\,\ref{fig:Compare_mom_LCDA}. The dashed lines represent the uncertainty associated with the initial resummation scale $\mu_0$, we systematically vary the scale within a range from $0.8\mu_0$ to $1.2\mu_0$ to assess the robustness of the perturbative matching procedure and to estimate the pertinent systematic uncertainty. 

\begin{figure}[!th]
\centering
\includegraphics[width=0.45\textwidth]{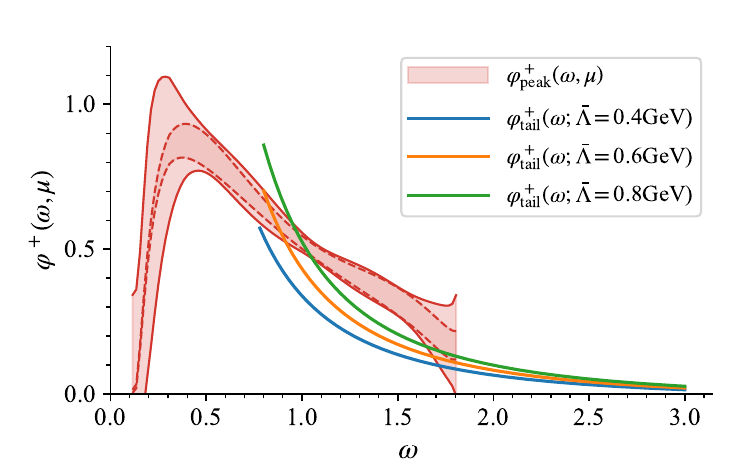}
\caption{The result of HQET LCDA in the peak region  and tail region. The dashed lines in $\varphi_{\mathrm{peak}}^+(\omega,\mu)$ indicate the upper and lower limits of the result containing only statistical error, while the solid lines indicate the result including both statistical and systematic errors.  }
\label{fig:HQET_LCDA_and_tail}
\end{figure}

The QCD LCDA can be divided into two distinct regions based on the order of the light quark momentum fraction $y$. When $y\sim\Lambda_{\mathrm{QCD}}/m_H$, corresponding to the region where a pronounced peak appears in $\phi(y,\mu)$. As discussed above, it can be factorized into the HQET LCDA $\varphi^+(\omega, \mu)$ at leading power of $\mathcal{O}(\Lambda_{\mathrm{QCD}}/m_H)$.  In the region $y\sim\mathcal{O}(1)$ which  is known as the ``tail region'', the QCD LCDA   contributes starting from the one-loop order in perturbation theory, and it perturbative result was firstly provided in Eq.(16) of Ref.\,\cite{Lee:2005gza}. In Fig.\,\ref{fig:HQET_LCDA_and_tail} we exhibit the results of $\varphi_{\mathrm{tail}}^+(\omega, \mu;\bar{\Lambda})$, where the parameter $\bar{\Lambda}$ reflects the power correction. 

Based on the normalized result for QCD LCDA provided above, we match the multiplicative factorization form given in Eq.(\ref{eq:peak_varphi_+}) to obtain the numerical result of HQET LCDA in the peak region. The result is shown by the pink band in Fig.\,\ref{fig:HQET_LCDA_and_tail}. As mentioned above, we include both the statistical error (shown as the dashed lines) and the systematic ones arise from data analysis. The solid lines of $\varphi_{\mathrm{peak}}^+$ represent the result incorporating both statistical and systematic errors.

\begin{figure}[!th]
\centering
\includegraphics[width=0.45\textwidth]{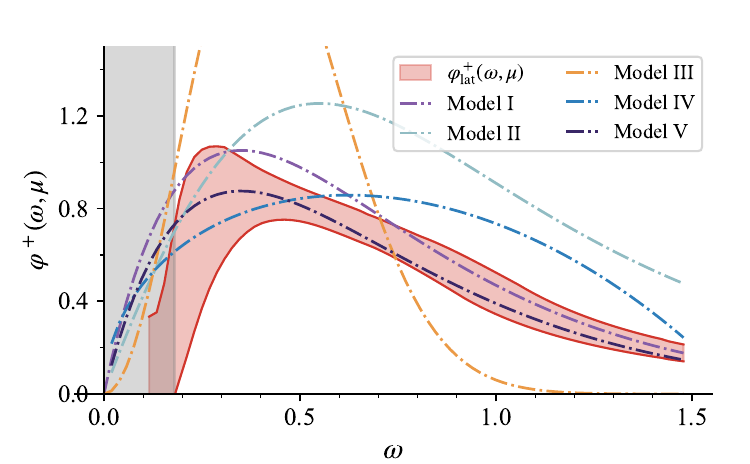}
\caption{Comparison of   numerical results of $\varphi^{+}\left(\omega, \mu\right)$ from lattice QCD with   commonly-used phenomenological models. The renormalization scale $\mu$ of our result is chosen to be $m_D$.  We only show the central values of the models as colored dashed lines, and interpret their spread as error estimate.}
\label{fig:comparetopheno}
\end{figure}

The power counting of parameters in HQET LCDA reveals that the momentum of the light quark in the peak region satisfies $\omega\sim\mathcal{O}(\Lambda_{\mathrm{QCD}})$, while in the tail region it satisfies $\omega\sim\mathcal{O}(m_D)$. This suggests that the boundary $\omega_b$ between them lies in the range $\Lambda_{\mathrm{QCD}}\ll\omega_{b}\ll m_D$. From Fig.\,\ref{fig:HQET_LCDA_and_tail}, we observe that the intersection of the peak and tail regions occurs around 0.9 GeV, confirming the aforementioned statement. Consequently, we merge the peak and tail regions and smooth the vicinity around the intersection to finally obtain a continuous distribution, as illustrated in Fig.\,\ref{fig:comparetopheno}. 
The grey band highlights the range where our findings are notably affected by power corrections, specifically around $\omega\lesssim0.19$ GeV.

In Fig.~\ref{fig:comparetopheno}, we compare our  results with the parametrizations for the HQET LCDA  based on  different models \cite{Grozin:1996pq,Braun:2003wx,Beneke:2018wjp}. The default parameters of the first four models are collected in Eq.(93) of Ref.\cite{Wang:2015vgv}, and of model V are given in Ref.~\cite{Beneke:2018wjp}.  Due to the challenges in estimating uncertainties in model-based analyses, we only display the central values as colored dashed lines. The spread of these lines serves as an estimate of systematic  errors in the models.  Our results do not favor the model II, III, and IV with their default parameters given in Ref.~\cite{Beneke:2018wjp}.  From the figure one can see that  the uncertainties are sizable, but future lattice simulations following similar methodologies can significantly improve precision, possibly eliminating the necessity for model-based analyses.

{\it Phenomenological  impacts: }The first inverse moment $\lambda_B^{-1}$ and inverse-logarithmic moments $\sigma_B^{(n)}$ play a pivotal role in QCD factorization theorems and  LCSR studies in heavy flavor physics.
Although the $\lambda_B^{-1}$ and  $\sigma_B^{(1)}$ have been studied in various approaches~\cite{Belle:2018jqd,Khodjamirian:2020hob,Lee:2005gza,Braun:2003wx,Grozin:1996pq}, there is considerable scope for enhancing their reliability and precision. In the case of $\sigma^{(2)}_B$ and subsequent orders, no existing results are currently available.

\begin{table}[htbp]
\caption{Upper panel: Numerical results for $\lambda_B$ and $\sigma_B^{(1)}$ at $\mu=1$GeV, which is consistent with the scale of theoretical results pesented in the lower panel.  Lower panel: $\lambda_B$ and $\sigma_B^{(1)}$ from experiment and other theoretical approaches. 
}
  \renewcommand{\arraystretch}{2.0}
  \setlength{\tabcolsep}{5.5mm}
  \begin{tabular}{ c c c }
    \hline\hline
      & $\lambda_B$ (GeV) & $\sigma_B^{(1)}$  \\
    \hline
     This work & 0.376(63) & 1.66(13)   \\
    \hline  
    Ref.\cite{Belle:2018jqd} & $>0.24$ &  \\
    Ref.\cite{Khodjamirian:2020hob} & $0.383(153)$ &  \\
    Ref.\cite{Lee:2005gza} & $0.48(11)$ & $1.6(2)$  \\
    Ref.\cite{Braun:2003wx} & $0.46(11)$ & $1.4(4)$  \\
    Ref.\cite{Grozin:1996pq} & $0.35(15)$ &  \\
    Ref.~\cite{Gao:2019lta} & $0.343^{+0.064}_{-0.079}$ &  \\
    Ref.\cite{Mandal:2023lhp} & $0.338(68)$ &  \\
    \hline
  \end{tabular} 
  \label{tab:model_independent_moments}
\end{table}

In the work, we employ model-independent parameterizations to extrapolate our results towards the region at small $\omega$, in order to reconstruct the fully distribution of HQET LCDA. 
{  One  feasible approach to achieve it  is to expand $\varphi^+(\omega, \mu)$ in the neighborhood where $\omega\to0$:
\begin{align}
\varphi^{+}&\left(\omega, \mu\right)=\sum_{n=1}^{N} c_n\frac{\omega^n}{\omega_0^{n+1}} e^{-\omega / \omega_0},
\label{eq:HQETLCDAsmallomegaexpansion}
\end{align}
The advantage of this parametrization is that, for small values of $\omega \ll \omega_0$, the influence of higher-order terms in the expansion become less influential. Another form based on the expansion of generalized Laguerre polynomials, recently proposed by Ref. \cite{Feldmann:2022uok}, is
\begin{align}
	\varphi^+(\omega,\mu)= \frac{\omega e^{-\omega / \omega_0}}{\omega_0^2} \sum_{k=0}^K \frac{a_k(\mu)}{1+k}L_k^{(1)}(2\omega/\omega_0), \label{eq:HQETLCDAsmallomegaexpansion2}
\end{align}
where $L_k^{(1)}$ denote the associated Laguerre polynomials. The advantage of this parametrization lies in its expansion based on a complete set of orthogonal polynomials.}
 
{ In the supplemental material \cite{supplemental}, we detail the utilization of two distinct parametrizations of HQET LCDAs,  along with expansions carried out up to the third order. Based on the extrapolated data, we determine the numerical values of  $\lambda_B$ and $\sigma_B^{(1)}$ of the HQET LCDA, as collected in the upper panel of Tab.\ref{tab:model_independent_moments}.  The errors reported for \(\lambda_B\) and \(\sigma_B^{(1)}\) in Table~\ref{tab:model_independent_moments} originate from a combination of statistical uncertainties and systematic variations introduced by differing parametrizations and expansion orders.  }

A comparison of these results with experimental data constraint~\cite{Belle:2018jqd} and phenomenological results~\cite{Khodjamirian:2020hob,Braun:2003wx,Grozin:1996pq} is given in the appendix, which indicate our findings for $\lambda_B$  are in agreement with them. While the results for $\sigma_B^{(1)}$  are only available from certain theoretical studies,  our results are still in approximate agreement with them.

To investigate the impact of the obtained $\lambda_B$ on the current studies of $B$ meson weak decays, we use the  results from LCSR~\cite{Gao:2019lta} to investigate the $B\to K^*$ form factors. 
As an example, we present the results for $\mathcal{V}(q^2=0)$. 
Compared to replacing the $\lambda_B$ used in Ref.\,\cite{Gao:2019lta} with our results, the relevant error estimates presented in Eq.\,(\ref{eq:ErrorsOfBtoKFF}) are refined as
\begin{align}
	\mathcal{V}_{B\to K^*}&(0)=0.359^{+0.141}_{-0.085}\Big|_{\lambda_B}{}^{+0.153}_{-0.079}\Big|_{\varphi_{\pm}}\pm\cdots \nn\\
	\Rightarrow \quad& \mathcal{V}_{B\to K^*}(0) = 0.297 \pm 0.088\Big|_{\lambda_B} \pm\cdots, 
\end{align}
in which we only focus on the terms originating from  heavy meson LCDA.  Central values of the form factors are smaller by $20\%$.  Specifically, reduced form factors may offer a potential solution to the issue of phenomenological  calculations surpassing experimental observations on differential branching fractions for $B\to K^*\ell^+\ell^-$~\cite{LHCb:2014cxe,LHCb:2016ykl}. 

{\it Summary and prospect: }
The first-principles calculation of HQET LCDAs for a heavy meson has posed a persistent challenge ever since the concept was introduced in 1990s~\cite{Grozin:1996pq}, with no successful resolution achieved thus far. 
In this Letter, we have presented a first-principle method for calculating the leading-twist HQET LCDA of heavy mesons.  To cure all the difficulties in the calculation, we have used a two-step matching method. In particular, we have constructed lattice QCD computable quasi-DAs characterized by three distinct dynamical scales  arranged in a hierarchical order $P^z\gg m_H\gg \Lambda_{\rm QCD}$. We have demonstrated that the separation of various dynamical scales can be achieved sequentially through a two-step factorization process, and the final outcome is  the heavy meson   LCDA.

To validate this approach, we have conducted a lattice QCD simulation of $D$ meson quasi-DA  and successfully matched it to the HQET LCDA. Our preliminary results exhibit qualitative agreement with several established phenomenological models. We have also extracted the first inverse moment, which lies in the experimentally constrained region derived from $B \to \gamma \ell\nu_\ell$ measurement and the phenomenological estimates.  More results can be found in a detailed supplement~\cite{LatticeParton:2024zko}. Following the same spirit  the heavy quark spin symmetry has been used to construct more useful matrix elements~\cite{Deng:2024dkd}, and an estimate of target mass corrections and renormalon ambiguities are conducted in Ref.~\cite{Han:2024cht}.  Dependence on heavy quark mass of QCD LCDAs is investigated in Ref.~\cite{Wang:2024wwa}.

This study marks the first instance in the literature of predicting the HQET LCDAs from lattice QCD.   After addressing various systematic effects,  it can be anticipated  that our methodology will  deliver reliable and  accurate  predictions on  LCDAs from first-principles of QCD, and facilitate  precise calculations for decay rates of heavy meson. 

{\it Acknowledgements:}
We thank Ding-Yu Shao,  Yu-Ming Wang, Yan-Bing Wei and LPC members for  valuable  discussions. We thank Yuming Wang for providing the results for the error budget in $B\to \pi$ form factors and his code for $B\to K^*$ form factors that allow to investigate the impact of our results on $B\to K^*$ form factors.  We thank C.-J. David Lin,  Andreas Sch\"afer, Yibo Yang, and particularly  Stefan Meinel for fruitful discussions on the simulation and the difficulties of HQET field on the lattice and drawing our attentions to \cite{Mandula:1990fit,Mandula:1993sj,Meinel:2010uji}.   We thank Jin-Xin Tan for an implementation of HQET field $h_v$ on the lattice which shows the signal-to-noise problem directly. We thank Thorsten Feldman on discussions on the model-independent parametrization of heavy meson LCDA. 
We also thank Ji-bo He on the discrepancies between experimental data and theoretical predictions regarding the differential branching fractions for $B\to K^*\ell^+\ell^-$ and the possible solution.   This work is supported in part by Natural Science Foundation of China under grant No. 12125503,   12335003, 12375069, 12105247, 12275277. Q.A.Z is also supported by the Fundamental Research Funds for the Central Universities. We thank the CLQCD collaborations for providing us the gauge configurations with dynamical fermions~\cite{Hu:2023jet}, which are generated on the HPC Cluster of ITP-CAS, the Southern Nuclear Science Computing Center(SNSC), the Siyuan-1 cluster supported by the Center for High Performance Computing at Shanghai Jiao Tong University and the Dongjiang Yuan Intelligent Computing Center. The computations in this paper were run on the Siyuan-1 cluster, and Advanced Computing East China Sub-center. The LQCD simulations were performed using the Chroma software suite~\cite{Edwards:2004sx} and QUDA~\cite{Clark:2009wm,Babich:2011np,Clark:2016rdz} through HIP programming model~\cite{Bi:2020wpt}. This work was partially supported by SJTU Kunpeng\&Ascend Center of Excellence.

\clearpage

\begin{widetext}

\section{SUPPLEMENTAL MATERIAL}

\subsection{A. Determination of $\lambda_B$ and $\sigma_B^{(1)}$}
To obtain the numerical results of $\lambda_B^{-1}$ and $\sigma_B^{(n)}$ from the values of the HQET LCDA we have developed, it is essential to perform the integrations:
\begin{align}
  \lambda_B^{-1}(\mu)=&\int_0^{\infty} \frac{d\omega}{\omega}  \varphi^+(\omega,\mu) , \nn\\
  \sigma^{(n)}_B(\mu) =& \lambda_B(\mu)\int_0^\infty \frac{d\omega}{\omega} \ln \left(\frac{\mu}{\omega}\right)^{(n)}\varphi^+(\omega,\mu). \label{eq:definitionofHQETmoments}
\end{align} 
which integrate $\varphi^+(\omega, \mu)$ over $\omega$ from 0 to infinity. As noted earlier, our results will be influenced by significant power corrections at small $\omega$, presenting difficulties for accurate predictions in this regime.
Alternatively, we can employ some model-independent parametrization formulas to extrapolate our results towards the region near $\omega=0$. A feasible approach is to expand $\varphi^+(\omega, \mu)$ in the neighborhood where $\omega\to0$:
\begin{align}
\varphi^{+}&\left(\omega, \mu\right)=\sum_{n=1}^{N} c_n\frac{\omega^n}{\omega_0^{n+1}} e^{-\omega / \omega_0},
\label{eq:HQETLCDAsmallomegaexpansion}
\end{align}
The advantage of this parametrization is that, for small values of $\omega \ll \omega_0$, the influence of higher-order terms in the expansion become less influential. Another form based on the expansion of generalized Laguerre polynomials, recently proposed by Ref. \cite{Feldmann:2022uok}, is
\begin{align}
	\varphi^+(\omega,\mu)= \frac{\omega e^{-\omega / \omega_0}}{\omega_0^2} \sum_{k=0}^K \frac{a_k(\mu)}{1+k}L_k^{(1)}(2\omega/\omega_0), \label{eq:HQETLCDAsmallomegaexpansion2}
\end{align}
where $L_k^{(1)}$ denote the associated Laguerre polynomials. The advantage of this parametrization lies in its expansion based on a complete set of orthogonal polynomials.

\begin{table*}[htbp]
\caption{Numerical results for $\lambda_B$ and $\sigma_B^{(1,2)}$ at $\mu=m_D$ and $\mu=1$ GeV, which are obtained from fits based on Eq.(\ref{eq:HQETLCDAsmallomegaexpansion}) up to the $N$-th order (labeled as ``Strategy I''), and from fits based on Eq.(\ref{eq:HQETLCDAsmallomegaexpansion2}) up to the $K$-th order (labeled as ``Strategy II''). The parameters of each fitting are appended subsequently.
}
  \renewcommand{\arraystretch}{2.0}
  \setlength{\tabcolsep}{3.5mm}
  \begin{tabular}{r | c c c | l }
    \hline\hline
        Order of Expansion  & $\lambda_B$ (GeV) & $\sigma_B^{(1)}$ & $\sigma_B^{(2)}$ & ~~~~~~~~~~~~~~~~~~~~~~~Pamameters  \\
    \hline
  $\mu=m_D$  ~~~    $N=1$ & 0.424(41) & 2.17(12) & 6.36(51) & $\omega_0=0.388 (46),~ c_1=0.916 (81)$ \\
   Strategy I:~~~  $N=2$ & 0.428(42) & 2.16(10)  & 6.29(47) &  $\omega_0=0.340(80),~ c_1=0.68(36), ~ c_2=0.11(16)$ \\
     $N=3$ & 0.418(67) & 2.18(15) & 6.33(80) & $\omega_0=0.32(15),~ c_1=0.63(44), ~  c_2=0.08(23), c_3=0.02(11)$ \\
   \hline 
    $\mu=m_D$  ~~~   $K=0$ & 0.421(33) & 2.17(8) & 6.35(32) & $\omega_0=0.389 (29), ~ a_0=0.925 (62)$  \\
    Strategy II:~~~ $K=1$ & 0.424(35) & 2.18(8) & 6.36(32) & $\omega_0=0.446 (95), ~ a_0=1.05 (19),~ a_1=0.15 (26)$  \\
     $K=2$ & 0.396(45) & 2.12(12) & 6.27(41) & $\omega_0=0.47 (10), ~ a_0=1.14 (21), ~ a_1=0.17 (24), ~ a_2=0.16 (18)$  \\
    \hline  
        \hline
    $\mu=1$ GeV  ~~~   $N=1$ &  0.380(37)  & 1.66(9)   &  \\
   Strategy I:~~~  $N=2$ &  0.384(38) & 1.65(8) &\\
     $N=3$   & 0.374(60) & 1.66(12) & \\
   \hline  
 $\mu=1$ GeV  ~~~    $K=0$ &  0.377(30) & 1.66(6)  &  \\
   Strategy II:~~~  $K=1$ &  0.380(31) & 1.67(6)  &\\
     $K=2$  & 0.356(41) & 1.62(9)  & \\
   \hline  
  \end{tabular} 
  \label{tab:inverse_moments}
\end{table*}

{In practice, we adopt the parameterization forms in both Eq.(\ref{eq:HQETLCDAsmallomegaexpansion}) and Eq.(\ref{eq:HQETLCDAsmallomegaexpansion2}) to extrapolate the lattice data to the small $\omega$ region. To avoid the influence of power corrections, we used the data at $\omega\in[0.192, 0.806]$GeV for fitting. 
Based on the fitted parameters, one can reconstruct the full distribution of the HQET LCDA and subsequently perform the integrals in Eq.(\ref{eq:definitionofHQETmoments}) to determine the numerical values of  $\lambda_B$ and $\sigma_B^{(n)}$. The values of fitted parameters and results for $\lambda_B$ and $\sigma_B^{(1,2)}$ are collected in Tab. \ref{tab:inverse_moments}. By integrating these results from various parameterizations and expansions to different orders, we obtain the corresponding constraints for  $\lambda_B$ and $\sigma_B^{(1,2)}$ at $\mu=m_D$:
\begin{align}
    &\lambda_B(\mu=m_D)  = 0.420(71)~ \mathrm{GeV}, \nn\\
    &\sigma_B^{(1)}(\mu=m_D)  = 2.17(16), \nn\\
    &\sigma_B^{(2)}(\mu=m_D)  = 6.33(80).
\end{align}
}

It should be noted that the HQET LCDA obtained from lattice QCD calculations, as well as the results for $\lambda_B$ and $\sigma_B^{(1,2)}$ derived from it, are obtained at $\mu=m_D$. To evolve them to the soft scale $\mu=1$ GeV which is commonly used in phenomenological studies, we apply the RG evolution for $\lambda_B$ and $\sigma_B^{(1)}$ as presented in Refs.\,\cite{Bell:2008er,Bell:2013tfa,Wang:2015vgv}, resulting in 
\begin{align}
  	&\lambda_B(\mu=1~\mathrm{GeV}) =0.376(63)~ \mathrm{GeV}, \nn\\
	&\sigma_B^{(1)}(\mu=1~\mathrm{GeV})  = 1.66(13),
\end{align}
where the errors are from the combinations of these individual errors.  
\end{widetext}

\end{document}